\documentclass[pra,twocolumn,showpacs,eqsecnum,superscriptaddress]{revtex4-1}

\usepackage{graphicx}
\usepackage{color}
\usepackage{amsmath}
\usepackage{microtype}

\usepackage{bm}
\usepackage{amsfonts}
\usepackage{amsbsy}

\begin{document}

\newcommand{\bra}[1]{\left\langle #1\right|}
\newcommand{\ket}[1]{\left|#1\right\rangle}
\newcommand{\braket}[2]{\left\langle #1|#2\right\rangle}
\newcommand{\com}[2]{\left[#1,#2\right]}
\newcommand{\braketop}[3]{\left\langle #1\left|#2\right|#3\right\rangle}
\newcommand{\mean}[1]{\left\langle #1 \right\rangle}
\newcommand{\trace}[2][]{{\rm Tr_{#1}}\left(#2\right)}
\newcommand{\ImaginaryPart}{{\rm Im}}
\newcommand{\RealPart}{{\rm Re}}
\newcommand{\leftexp}[2]{{\vphantom{#2}}^{#1}{#2}}
\newcommand{\leftind}[2]{{\vphantom{#2}}_{#1}{#2}}
\newcommand{\elem}{\in}
\newcommand{\rp}{\right)}
\newcommand{\lp}{\left(}
\newcommand{\lcb}{\left\{}
\newcommand{\rcb}{\right\}}
\newcommand{\rsb}{\right]}
\newcommand{\lsb}{\left[}
\newcommand{\lbv}{\left|}
\newcommand{\rbv}{\right|}
\newcommand{\lvb}{\lbv}
\newcommand{\rvb}{\rbv}
\newcommand{\bs}{\boldsymbol}
\renewcommand{\inf}{\infty}
\newcommand{\myfrac}[2]{^{#1\negthickspace\negthickspace}/_{\negthinspace#2}}
\newcommand{\mycaption}[2]{\caption[#1]{\small #1 #2}}
\newcommand{\order}[1]{{{\mathcal O}\lp#1\rp}}
\newcommand{\iohbar}{\frac{-i}{\hbar}}
\newcommand{\melem}[1]{_{#1}}
\newcommand{\pref}[1]{(\ref{#1})}
\renewcommand{\eqref}[1]{Eq.~\pref{#1}}
\newcommand{\hc}{\mathrm{h.c.}}

\newcommand{\qrho}{\rho}
\newcommand{\crho}{\varrho}
\newcommand{\gd}{\gamma_\downarrow}
\newcommand{\gu}{\gamma_\uparrow}

\newcommand{\varlambda}{\Lambda}
\newcommand{\varchi}{X}
\newcommand{\varS}{\mathbb{S}}
\newcommand{\varK}{\mathbb{K}}

\newcommand{\superop}[1]{{\mathcal #1}}
\newcommand{\sD}{{\superop{D}}}
\newcommand{\sL}{{\superop{L}}}
\newcommand{\sC}{{\superop{C}}}
\newcommand{\sT}{{\superop{T}}}
\newcommand{\sM}{{\superop{M}}}
\newcommand{\sB}{{\superop{B}}}

\newcommand{\tr}[1]{\mathbf{#1}}
\newcommand{\tU}{{\tr{U}}}
\newcommand{\tS}{{\tr{S}}}
\newcommand{\tD}{{\tr{D}}}
\newcommand{\tR}{{\tr{R}}}
\newcommand{\tP}{{\tr{P}}}
\newcommand{\tT}{{\tr{T}}}
\newcommand{\trans}[1]{^{#1}}

\newcommand{\atr}[1]{{\mathbf #1}}
\newcommand{\atU}{{\atr{U}}}
\newcommand{\atS}{{\atr{S}}}
\newcommand{\atD}{{\atr{D}}}
\newcommand{\atR}{{\atr{R}}}
\newcommand{\atP}{{\atr{P}}}
\newcommand{\atT}{{\atr{T}}}

\newcommand{\g}{g}
\newcommand{\e}{e}
\newcommand{\f}{f}

\newcommand{\wg}{\omega_\g}
\newcommand{\we}{\omega_\e}
\newcommand{\wf}{\omega_\f}
\renewcommand{\wr}{\omega_r}
\newcommand{\wm}{\omega_d}

\newcommand{\proj}[1]{\Pi_{#1}}

\newcommand{\ad}{a^\dag}
\newcommand{\ada}{a^\dag a}
\newcommand{\sz}{\sigma_0}
\newcommand{\szd}{\sigma_0^\dag}
\newcommand{\so}{\sigma_1}
\newcommand{\sod}{\sigma_1^\dag}
\newcommand{\gz}{g_0}
\newcommand{\go}{g_1}
\newcommand{\izp}{I_{0+}}
\newcommand{\izm}{I_{0-}}
\newcommand{\izpm}{I_{0\pm}}
\newcommand{\izmp}{I_{0\mp}}
\newcommand{\iop}{I_{1+}}
\newcommand{\iom}{I_{1-}}
\newcommand{\iopm}{I_{1\pm}}
\newcommand{\iomp}{I_{1\mp}}
\newcommand{\itp}{I_{2+}}

\newcommand{\epm}{\epsilon_m}
\newcommand{\Lz}{\Lambda_0}
\newcommand{\lz}{\lambda_0}
\newcommand{\Dz}{\Delta_0}
\newcommand{\Zz}{Z_0}
\newcommand{\chiz}{\chi_0}
\newcommand{\zetaz}{\zeta_0}
\newcommand{\xiz}{\xi_0}
\newcommand{\Lo}{\Lambda_1}
\newcommand{\lo}{\lambda_1}
\newcommand{\Do}{\Delta_1}
\newcommand{\Zo}{Z_1}
\newcommand{\chio}{\chi_1}
\newcommand{\zetao}{\zeta_1}
\newcommand{\xio}{\xi_1}

\newcommand{\subge}{\mathcal{E}_{\g\e}}
\newcommand{\sm}{\sigma_-}
\renewcommand{\sp}{\sigma_+}
\newcommand{\sigmaz}{\sigma_z}
\newcommand{\comm}[2]{\lsb #1,#2\rsb}

\newcommand{\red}{\color[rgb]{0.8,0,0}}
\newcommand{\green}{\color[rgb]{0.0,0.6,0.0}}
\newcommand{\dkgrn}{\color[rgb]{0.0,0.4,0.0}} 
\newcommand{\blu}{\color[rgb]{0,0,0.6}}
\newcommand{\blue}{\color[rgb]{0,0,0.6}}
\newcommand{\pur}{\color[rgb]{0.8,0,0.8}}
\newcommand{\blk}{\color{black}}

\newcommand{\ab}[1]{{\red AB:~#1}} 

\title{Straddling regime of circuit QED to improve bifurcation readout}
\title{Improved qubit bifurcation readout in the straddling regime of circuit QED}
\date{\today}

\author{Maxime~Boissonneault}
 \affiliation{D\'epartement de Physique, Universit\'e de Sherbrooke, Sherbrooke, Qu\'ebec, Canada, J1K 2R1}
\author{J.~M.~Gambetta}
\affiliation{IBM T.J. Watson Research Center, Yorktown Heights, NY 10598, USA}
\author{A.~Blais}
 \affiliation{D\'epartement de Physique, Universit\'e de Sherbrooke, Sherbrooke, Qu\'ebec, Canada, J1K 2R1}

\begin{abstract}
We study bifurcation measurement of a multi-level superconducting qubit using a nonlinear resonator biased in the straddling regime, where the resonator frequency sits between two qubit transition frequencies. We find that high-fidelity bifurcation measurements are possible because of the enhanced qubit-state-dependent pull of the resonator frequency, the behavior of qubit-induced nonlinearities and the reduced Purcell decay rate of the qubit that can be realized in this regime. Numerical simulations find up to a threefold improvement in qubit readout fidelity when operating in, rather than outside of,  the straddling regime. High-fidelity measurements can be obtained at much smaller qubit-resonator couplings than current typical experimental realizations,  reducing spectral crowding and potentially simplifying the implementation of multi-qubit devices.
\end{abstract}


\maketitle

\section{Introduction} 
\label{sec:introduction}
Circuit quantum electrodynamics (cQED), where superconducting qubits are coupled to transmission-line resonators, constitute a promising architecture for the realization of a quantum information processor~\cite{Blais2004,Wallraff2004}. 
Two criteria required for quantum computation are the implementation, in a scalable way, of a universal set of gates and the ability to faithfully measure the qubit state~\cite{DiVincenzo2000}. In this system, single qubit gates can be performed by sending microwave signals through the resonator close to the qubits' transition frequency, while two-qubit gates can be performed by tuning the qubits in and out of resonance. The increasing fidelity of one-~\cite{Wallraff2005} and two-qubit~\cite{Majer2007,Niskanen2007,Chow2011} gates has allowed circuit QED to reach important milestones, such as the implementation of two- and three-qubit quantum algorithms~\cite{DiCarlo2009,Fedorov2012,Reed2012} and the realization of more complex multi-qubit devices~\cite{Mariantoni2011a}.

Qubit measurement in cQED is realized by driving the resonator close to its natural resonance frequency and by measuring the reflected or transmitted microwave signal. Recently, high-fidelity single-shot measurements have been achieved by using very large measurement drive powers~\cite{Reed2010,Boissonneault2010,Bishop2010}, by turning the resonator into a nonlinear active device and using bifurcation to distinguish the qubit states~\cite{Siddiqi2004,Lupascu2006,Mallet2009}, or by using nearly quantum-limited amplifiers~\cite{Vijay2011}. In these realizations, increasing the qubit-resonator coupling leads to larger variation of the resonator's parameters with the qubit state, resulting in high measurement fidelity. In the same way, increasing this coupling also typically reduces the gate-time of two-qubit operations. However, stronger coupling can also reduce the on/off ratio of logical gates, causes spectral crowding and reduces the qubit lifetime through spontaneous emission via the resonator, also known as the Purcell effect.

In this paper, we take a different approach and show that it is possible to implement high-fidelity single-shot measurements of a superconducting qubit using relatively small qubit-resonator coupling strengths --- of the order of $10$~MHz --- than in many recent experiments. To achieve this, we use the weakly anharmonic multi-level structure relevant for most superconducting qubits and take advantage of the so-called straddling regime where the resonator frequency sits between two qubit transitions~\cite{Koch2007}. This regime shows enhanced qubit-state-dependent pull of the resonator frequency, enhanced qubit-induced resonators and reduced Purcell decay rate. We show that these three characteristics combine to improve bifurcation measurements of the qubit state. In numerical simulations of qubit readout, we find error probabilities three times smaller inside with respect to outside of the straddling regime. Even without thorough exploration of the available parameter space, we find measurement fidelities of 98\%.

The paper is organized as follows. In section~\ref{sec:initial_model}, we first introduce the Hamiltonians modeling a nonlinear resonator, required for bifurcating measurements, coupled to a multi-level qubit. Then, in section~\ref{sec:the_principle_of_bifurcation_measurements}, we review the principle of bifurcation measurements and highlight the important differences between two-level and multi-level qubits in this respect. In section~\ref{sec:dispersive_model_in_the_straddling_regime}, we derive an effective dispersive Hamiltonian valid in the straddling regime. Finally, we compare in section~\ref{sec:improving_bifurcation_measurements_with_the_straddling_regime} the parameters calculated with our model to parameters extracted from exact diagonalization of the qubit-resonator Hamiltonian. We then examine the specifics of bifurcation in the straddling regime, extract measurement fidelities from numerical simulations and discuss other advantages of working in this regime. 

\section{Model} 
\label{sec:initial_model}
As mentioned above, many superconducting qubits have a relatively small anharmonicity and are therefore described by M-level systems with $M>2$ rather than by two-level systems~\cite{Koch2007,Schreier2008,Houck2009,Steffen2010}. We consider such a qubit coupled to a Kerr nonlinear resonator, which could be realized for example by an LC-circuit with a Josephson junction~\cite{Siddiqi2004} or a stripline resonator with one~\cite{Mallet2009} or many~\cite{Castellanos-Beltran2007,Yamamoto2008} embedded Josephson junctions making it nonlinear. The qubit-resonator  system can be modeled with the many-level version of the Jaynes-Cummings Hamiltonian
\begin{equation}
	\label{eqn:H_s}
	H_s = H_q + H_r + H_I,
\end{equation}
where ($\hbar=1$)
\begin{equation}
	\label{eqn:H_q}
	H_q = \sum_{i=0}^{M-1} \omega_i \proj{i,i} \equiv \proj{\omega},
\end{equation}
is the qubit Hamiltonian, 
\begin{equation}
	\label{eqn:H_r}
	H_r = \wr\ada + \frac{K}{2} \ad\ad aa,
\end{equation}
is the nonlinear resonator Hamiltonian~\cite{Yurke2006}, and 
\begin{equation}
	\label{eqn:H_I}
	H_I = \sum_{i=0}^{M-2} g_i (\ad\proj{i,i+1} + a\proj{i+1,i}),
\end{equation}
is the interaction Hamiltonian and where $\proj{i,j}\equiv\ket{i}\bra{j}$ with $\{\ket{i}\}$ the qubit eigenstates. In these expressions, $\omega_i$ is the frequency associated to the qubit eigenstate $\ket{i}$, $\wr$ is the bare resonator frequency (at low powers), $K$ is the Kerr constant, and $g_i$ the qubit-resonator coupling constants. We have also introduced the short-hand notation
\begin{equation}
	\label{eqn:proj_x}
	\proj{x} \equiv \sum_{i=0}^{M-1} x_i \proj{i,i},
\end{equation}
where $x$ is a scalar taking different values $x_i$ associated to the different qubit states $\ket i$. This notation is used  throughout this paper. Finally, in the qubit-resonator interaction term, we have made the standard rotating-wave approximation (RWA) and also assumed that transition between states $\ket{i}\leftrightarrow\ket{j}$ are suppressed for $|i-j|\ne 1$~\cite{Koch2007}.  

Measurement of the qubit is realized by driving the resonator with a tone of amplitude $\epsilon_d$ and frequency $\omega_d$. This is modeled by the drive Hamiltonian
\begin{equation}
	H_d = \epsilon_d (e^{-i\omega_d t}\ad + e^{i\omega_d t} a),
\end{equation}
leading to the total Hamiltonian
\begin{equation}
	\label{eqn:H}
	H = H_s + H_d.
\end{equation}

\section{Basics of bifurcation measurements} 
\label{sec:the_principle_of_bifurcation_measurements}
\begin{figure}
	\centering
	\includegraphics[width=0.95\hsize]{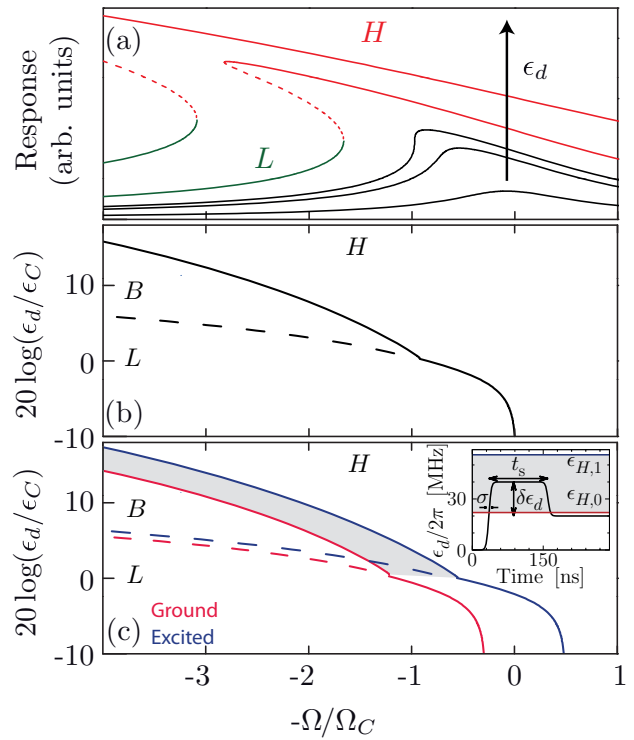}
	\caption{(Color online) (a) Response (amplitude of the field) as a function of the reduced detuning and for increasing drive amplitude $\epsilon_p$. The back-bending of the response reflects the choice $K<0$, as is usually the case in circuit QED (b,c) Stability diagram of a Kerr nonlinear resonator in the absence (b) or presence (c) of a qubit. Inset : Time-dependent enveloppe of a sample-and-hold readout.}
	\label{fig:bifurcation_phase_diagram_qubit}
\end{figure}

The description of the Kerr nonlinear resonator (KNR) is simplified by introducing the reduced detuning frequency $\Omega \equiv 2(\omega_r-\omega_d)/\kappa$~\cite{Ong2011}. As illustrated in Fig.~\ref{fig:bifurcation_phase_diagram_qubit}, the steady-state response of the KNR can vary drastically whether the reduced detuning $\Omega$ is larger or smaller (in absolute value) than a critical detuning $\Omega_C=\sqrt{3}$. For $|\Omega/\Omega_C| < 1$, the resonator response is single valued, with as shown in Fig.~\ref{fig:bifurcation_phase_diagram_qubit}~(a), a response that is stiffened compared to the usual Lorentzian line shape. Close, but below, the critical point, the resonator can then be used as a parametric amplifier for small signals~\cite{Castellanos-Beltran2007}. On the other hand, for $|\Omega/\Omega_C| > 1$ the resonator is in the so-called bifurcation amplification regime (BA) where it is bistable for a range of drive amplitudes $\epsilon_p$. If $\epsilon_p$ is ramped up starting from zero, the resonator's response will bifurcate from a low ($L$) to a high ($H$) oscillation amplitude dynamical state at a critical amplitude $\epsilon_H$. If the drive amplitude is then reduced, the resonator stays in the state $H$ until the drive amplitude becomes lower than a second threshold $\epsilon_L$. The associated stability diagram is illustrated in Fig.~\ref{fig:bifurcation_phase_diagram_qubit}~(b).

As was already experimentally demonstrated, in the BA regime, the KNR can be used as a sample-and-hold detector of a qubit~\cite{Siddiqi2006,Lupascu2006,Mallet2009,Vijay2009}. Indeed, as for most quantum information related tasks, qubit readout is realized in the dispersive regime where $|g_i| \ll |\omega_r-\omega_{i+1,i}|$. In this situation, the system Hamiltonian $H_s$ is well approximated by the effective Hamiltonian~\cite{Boissonneault2012b}
\begin{equation}
	H_D \approx (\omega_r+\proj{S})\ada + \frac{K}{2}\ad\ad aa + \proj{\omega}.
\end{equation}
As can be seen from the coefficient of $a^\dag a$, in this regime, the presence of the qubit results in a shift of the resonator frequency by a qubit-state dependent quantity $S_i$. This dispersive cavity pull, whose value $S_i$ will be discussed below, results in different thresholds $\epsilon_{L,i}$ and $\epsilon_{H,i}$ depending on the qubit states. This is schematized for the first two qubit states $\{\ket 0, \ket 1\}$ by the red and the blue lines in Fig.~\ref{fig:bifurcation_phase_diagram_qubit}~(c). 

Starting from zero, increasing the drive amplitude $\epsilon_d$ until $\epsilon_{H,0} < \epsilon_d < \epsilon_{H,1}$ will then result in a high amplitude of the cavity field if the qubit is in its ground state and a low amplitude if it is in its excited state. This range is represented by the gray shaded area in Fig.~\ref{fig:bifurcation_phase_diagram_qubit}~(c). If the drive amplitude is then reduced below $\epsilon_{H,0}$, but stays above $\epsilon_{L,1/0}$ [see inset of Fig.~\ref{fig:bifurcation_phase_diagram_qubit}~(c)], both resulting states are stable and the qubit state has been mapped into the dynamical state of the resonator. Since these dynamical state are stable, it is possible to accumulate the output signal for a time longer than the qubit relaxation time $T_1$. The measurement fidelity can then be optimized by varying the sampling time $t_s$, the height of the plateau $\delta\epsilon_m$ and the steepness of the ramp up $\sigma$~\cite{Siddiqi2006,Lupascu2006,Mallet2009,Vijay2009}. 

In practice, the readout fidelity is limited by qubit relaxation during or before the sample phase~\cite{Mallet2009}, when the resonator has not bifurcated yet. The speed at which the sampling can be made is limited by the resonator's decay rate $\kappa$. Indeed, ramping up the drive much faster than $1/\kappa$ will produce large ringing oscillations in the field amplitude and which can result in false positives or negatives. This results in a reduced measurement fidelity. Increasing $\kappa$ therefore implies smaller transients and hence faster measurement. However, increasing $\kappa$ too much can also yield a lower measurement fidelity. Indeed, in the limit where $\kappa$ is much larger than the difference between the qubit-state dependent resonator pulls $\chi\equiv S_1-S_0$, both qubit states are indisthinguishable. Moreover, increasing $\kappa$ also increases the qubit's Purcell decay rate $\gamma_\kappa \sim \kappa g_{i}^2/(\omega_{i+1}-\omega_i-\omega_r)^2$~\cite{Purcell1946,Houck2008} which ultimately limits the qubit relaxation time $T_1$. Ideally, one would like to increase both $\chi$ and $\kappa$, without increasing the Purcell rate.

\subsection{Two-level systems} 
\label{sub:two_level_systems}
The qubit-state dependent resonator shift $S_i$ discussed above depends on the coupling $g_i$ and the qubit-resonator detuning $\Delta_{i,r}\equiv\omega_{i+1}-\omega_i-\omega_r$. For a two-level qubit, it takes the simple form~\cite{Blais2004}
\begin{equation}
	S_{1/0}^{\rm 2LS} = \pm \frac{g_0^2}{\Delta_{0,r}},
\end{equation}
corresponding to symmetric displacement of the cavity frequency around its bare frequency $\omega_r$. The difference between the pulled resonator frequency for the qubit states $\ket{0}$ and $\ket{1}$ is therefore $\chi = S_{1} - S_{0} = 2g_0^2/\Delta_{0,r}$. This cavity pull can be of the order of a few tens of MHz, while staying in the dispersive regime, with the typical values $g_0/2\pi\sim 100-200$~MHz and $\Delta_{0,r}/2\pi\sim1-2$~GHz. Such couplings have been achieved with transmons and flux qubits ~\cite{Schuster2007,Abdumalikov2008}. For a two-level qubit, increasing the coupling $g_{0}$ increases $\chi$, but also increases $\gamma_\kappa$ by the same amount. For Purcell-limited qubits, this negates the gain of this strategy.

\subsection{Multi-level systems} 
\label{sub:multi_level_systems}
\begin{figure}
	\centering
	\includegraphics[width=0.95\hsize]{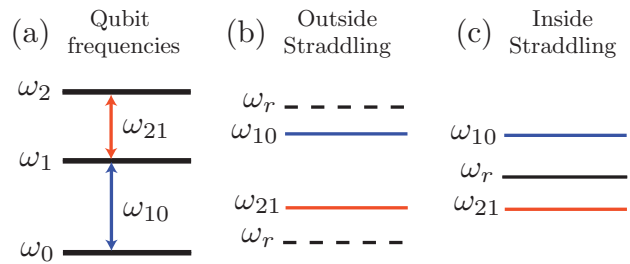}
	\caption{(Color online) (a) Absolute and transition frequencies of the first three eigenstates of a qubit with negative anharmonicity such as the transmon. (b) Example of frequency diagram where the resonator frequency (dashed lines) is outside of a straddling regime. (c) Example of energy diagram where the resonator frequency (full black line) is inside of a straddling regime. }
	\label{fig:frequency_diagram}
\end{figure}

For multi-level systems, the shifts are changed by the presence of additional levels, and the symetry around the bare resonator frequency is broken. Indeed, the frequency shift is given by~\cite{Koch2007}
\begin{equation}\label{eq:Smls}
	S_i^{\rm MLS} = \frac{g_{i-1}^2}{\Delta_{i-1,r}}-\frac{g_i^2}{\Delta_{i,r}}.
\end{equation}
As illustrated in Fig.~\ref{fig:frequency_diagram}~(b), in most experiments~\cite{DiCarlo2009,Mallet2009,Ong2011} the qubits are biased such that the resonator frequency sits above, or below, all of the qubit transition frequencies. This results in a pull of the resonator frequency $\chi = 2g_0^2/\Delta_{0,r} - g_1^2/\Delta_{1,r}$, reduced compared to that of a purely two-level system. In the limit where the multi-level system tends toward a harmonic oscillator, $\Delta_{1,r}\rightarrow\Delta_{0,r}$ and $g_1\rightarrow\sqrt{2}g_0$ such that this pull vanishes. The reduction in the pull can be compensated with larger couplings $g_i$ achieved for example with transmons~\cite{Schreier2008}. However, as stated above, increasing the qubit-resonator coupling also increases the resonator-mediated Purcell decay~\cite{Houck2008} and dressed dephasing~\cite{Boissonneault2008,Boissonneault2009,Wilson2010}. This dependence on $\kappa$ and $g_i$ of both resonator-mediated qubit decay and measurement speed ultimately limits the achievable measurement fidelity.

One way to increase the dispersive shifts $\chi$ without increasing the coupling is to work in the so-called straddling regime~\cite{Koch2007}. In this regime, illustrated in Fig.~\ref{fig:frequency_diagram}~(c), the detunings $\Delta_{1,r}$ and $\Delta_{0,r}$ are of opposite signs. As a result, instead of canceling each other, the two terms in \eqref{eq:Smls} add up, yielding a significantly enhanced value of $\chi$. Since this improvement is obtained without increasing $g_i$, it does not increase the Purcell rate. Moreover, as we will show in the next section, this regime also increases qubit-induced nonlinearities~\cite{Boissonneault2010}, something that we will exploit below to improve bifurcation readouts.

\section{Dispersive model in the straddling regime} 
\label{sec:dispersive_model_in_the_straddling_regime}
Following the approach of Ref.~\cite{Boissonneault2012b}, we use a polaron transformation~\cite{Mahan2000,Irish2005,Gambetta2008} followed by a dispersive transformation~\cite{Carbonaro1979,Boissonneault2008} to approximately diagonalize the Hamiltonian of \eqref{eqn:H}. Doing the transformations in this order (polaron followed by dispersive), allows to correctly model the ac-Stark shift caused by a drive detuned from the resonator frequency~\cite{Boissonneault2012b}. However, since we are interrested in the straddling regime, one more transformation must to be done in order to diagonalize an effective two-photon process that is important only in the straddling regime. This is done in Appendix~\ref{annsec:dispersive_transformation_of_the_two_photon_terms} and yields the effective diagonal Hamiltonian
\begin{equation}
	\label{eqn:H_fourth}
	H' = \proj{\omega'} + [\omega_r'(\alpha) + \proj{S(\alpha)}]\ada,
\end{equation}
where $\omega_r'(\bar\alpha)$ is the Kerr-shifted resonator frequency
\begin{equation}
	\omega_r'(\alpha) \equiv \omega_r+2K|\bar\alpha|^2,
\end{equation}
with $\bar\alpha\equiv \mean{\proj{\alpha}}$ the resonator mean field and
\begin{equation}
	\label{eqn:omega_fourth}
	\omega_i' = \omega_i + \varS_i |\bar\alpha|^2 + \varK_i |\bar\alpha|^4 + L_i(\bar\alpha), 
\end{equation}
are the renormalized effective qubit frequencies. There, we have defined 
\begin{subequations}
	\label{eqn:classical_stark_shift_coefficients}
	\begin{align}
		\label{eqn:varS_i}
		\varS_i &\equiv -(\varchi_i - \varchi_{i-1}), \\
		\label{eqn:varK_i}
		\varK_i &\equiv -\varS_i(|\varlambda_i|^2+|\varlambda_{i-1}|^2) \\
		&\quad - \tfrac14(3\varchi_{i+1}|\varlambda_i|^2-\varchi_i|\varlambda_{i+1}|^2) \nonumber \\
		&\quad + \tfrac14(3\varchi_{i-2}|\varlambda_{i-1}|^2-\varchi_{i-1}|\varlambda_{i-2}|^2) \nonumber \\
		&\quad - \varchi_{i}^{(2)} + \varchi_{i-2}^{(2)}, \nonumber
	\end{align}
\end{subequations}
the linear and quadratic ac-Stark shift coefficients with $\varlambda_i \equiv -g_i/\Delta_{i,d}$, $\varchi_i \equiv -g_i\varlambda_i$, and where $\Delta_{i,d} \equiv \omega_{i+1} - \omega_i - \omega_d$ is the detuning between the qubit transition $i$ and the drive $d$. The last line of $\varK_i$ comes from the diagonalization of an effective two-photon transition process that is large only in the straddling regime. This contributes the last two terms of $\varK_i$ with $\varchi_i^{(2)} = -g_i^{(2)} \varlambda_i^{(2)}$, $\varlambda_i^{(2)} = -g_i^{(2)}/(\Delta_{i+1,d}+\Delta_{i,d})$ and where
\begin{equation}
	\label{eqn:g_i_2}
	g_{i}^{(2)} = \varlambda_{i}\varlambda_{i+1} (\Delta_{i+1,d} - \Delta_{i,d}).
\end{equation}
We note that, when compared with the results of Ref.~\cite{Boissonneault2010}, the detunings $\Delta_i$ are defined with respect to the \emph{drive} frequency, and not the \emph{resonator} frequency. In addition, in Ref.~\cite{Boissonneault2010}, the dispersive transformation was done with respect to the field operator $a$ rather than to the classical field $\alpha$. Because of this choice, the quadratic term $\varS_i$ in Ref.~\cite{Boissonneault2010} contains correction which accounts for a specific choice of ordering for the ladder operators in the quartic term. Here, since it is the classical field that is considered, there are no such corrections (i.e. $\alpha \alpha^* = \alpha^*\alpha$).

Finally, in \eqref{eqn:omega_fourth} we have also defined the Lamb shift
\begin{equation}
	\label{eqn:L_i}
	L_i(\alpha) = \frac{g_i^2}{\omega''_{i+1}(\alpha)-\omega''_i(\alpha)-\omega_r'(\alpha)},
\end{equation}
where $\omega_i''$ is given by
\begin{equation}
	\omega_i'' = \omega_i + \varS_i |\bar\alpha|^2 + \varK_i^{(1)}|\bar\alpha|^4.
\end{equation}
Using this definition, the cavity pull $\proj{S(\alpha)}$ in the effective Hamiltonian \eqref{eqn:H_fourth} can be expressed in a compact way using $S_{i}(\alpha) = -[L_{i+1}(\alpha)-L_i(\alpha)]$. We note that while the ac-Stark shift coefficients $\varS_i$ and $\varK_i$ depend on the qubit-drive detuning, the Lamb-shift $L_i$ depends on the detuning between the ac-Stark shifted qubit and Kerr-shifted resonator~\cite{Boissonneault2012b}. Finally, the steady-state qubit-state-dependent cavity field $\alpha_i$ is given by the solution of  
\begin{equation}
	\label{eqn:condition_alpha_d}
	-\epsilon_d = (\omega_r-\omega_d-i\tfrac{\kappa}{2})\alpha_i + K|\alpha_i|^2\alpha_i + \lp\varS_i+\frac{4}{3!}\varK_i|\alpha_i|^2\rp \alpha_i.
\end{equation}
\begin{figure}
	\centering
	\includegraphics[width=0.95\hsize]{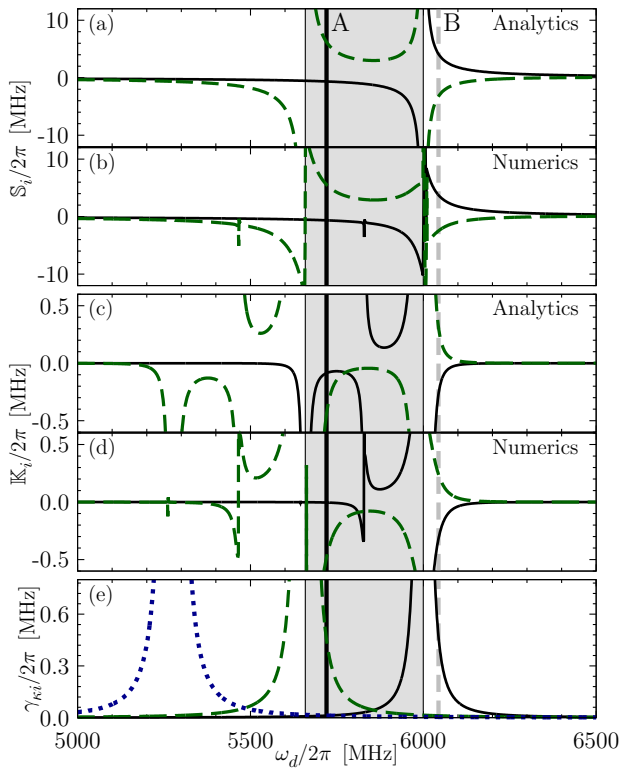}
	\caption{(Color online) [(a) and (b)] Linear and [(c) and (d)] quadratic ac-Stark shifts for the ground (full black lines) and the excited (green dashed lines) states of a transmon qubit~\cite{Koch2007} with charging energy $E_C=300$~MHz, Josephson energy $E_J=25$~GHz, coupling at zero flux $g_{10}/2\pi=15$~MHz tuned such that $\omega_{10}/2\pi=6$~GHz. This yields $(\omega_0,\omega_1,\omega_2,\omega_3)/2\pi\approx(0,6,11.7,16.9,21.8)$~GHz and $(g_{10},g_{21},g_{32},g_{43})/2\pi=(13.5,18.5,21.8,24.1)$~MHz at the operating point. Panels (a) and (c) correspond to the analytical equations~\eqref{eqn:classical_stark_shift_coefficients}, while panels (b) and (d) are extracted numerically as described in the text. Panel (e) shows the Purcell decay rate $\gamma_{\kappa i}=\kappa g_i^2/\Delta_{i,r}^2$ assuming $\kappa/2\pi=5$~MHz and $\omega_r=\omega_d$ for $i=0$ (full black line), $i=1$ (dashed green line) and $i=2$ (dotted blue line). The black line, corresponding to the $\ket 1 \rightarrow \ket 0$ transition, is the relevant one for qubit operation. The shaded area corresponds to the straddling regime. The full vertical black line and the dashed vertical grey line correspond to the two operating points A ($\omega_d/2\pi=5720$~MHz) and B ($\omega_d/2\pi=6044$~MHz) discussed in Secs.~\ref{sec:dispersive_model_in_the_straddling_regime} and \ref{sec:improving_bifurcation_measurements_with_the_straddling_regime}. }
	\label{fig:stark_kerr}
\end{figure}

In Fig.~\ref{fig:stark_kerr}, we compare the above analytical expressions for $\varS_i$ and $\varK_i$ to numerical results. These quantities are found numerically by fitting a quadratic polynomial to the resonator frequency for the qubit state $\ket i$ and in the presence of $n$ photons, $\omega_{r,i}(n) = E_{i,n+1}-E_{i,n}$. The energy $E_{i,n}$ is found numerically by diagonalizing the undriven qubit-resonator Hamiltonian $H_s$ and taking $K=0$. We then associate $E_{i,n}$ to the energy of the eigenstate closest to the bare qubit-resonator state $\ket{i,n}$. The parameters, given in the caption of Fig.~\ref{fig:stark_kerr}, are typical to transmon qubits~\cite{Koch2007}, but with a smaller than typical coupling $g_{10}/2\pi=13.5$~MHz. We show the analytical [(a) and (c)] and numerical [(b) and (d)] values of $\varS_i$ [(a) and (b)] and $\varK_i$ [(c) and (d)] for the ground state $i=0$ (full black lines) and first qubit excited states $i=1$ (dashed green lines). We find quantitative agreement, except at the qubit-resonator resonances and at the two-photon resonances (identified by divergences). We finally show in Fig.~\ref{fig:stark_kerr}~(e) the Purcell decay rate $\gamma_{\kappa i}$ of level $i$ assuming $\omega_r=\omega_d$ and $\kappa/2\pi=5$~MHz. 

Two operating points, designated by A and B and identified by the vertical full black lines and dashed grey lines respectively, are illustrated on Fig.~\ref{fig:stark_kerr}. These particular points have been chosen because, while A lies in the straddling regime and B is outside of that regime, the cavity pull $|\chi|=|\varS_1-\varS_0|$ is identical in both cases. In the next section, we will show that working in the straddling regime is advantageous for qubit readout. Since the cavity pull is the same at both $A$ and $B$,  improvement in the measurement will be due to qubit-induced nonlinearities $\varK_i$ or variation in the Purcell decay rate. 

The qubit-induced nonlinearities $\varK_i$ are plotted in panels (c) and (d). Comparing panels (a) and (c), we note a major difference between the operating points A and B. At B, the sign of $\varK_i$ is opposite to that of $\varS_i$ for both $i=0$ (full black lines) and $i=1$ (dashed green lines). This sign difference corresponds to a cavity pull that is \emph{decreasing} when the number of photons increases. On the other hand, at point A, the sign of $\varK_0$ is the same as that of $\varS_0$. Therefore, we expect that the cavity pull at point A will not decrease as much as at point B with increasing photon number~\cite{Boissonneault2010}. Moreover, we can see in panel (e) that the Purcell rate for the transition $\ket{1}\rightarrow\ket{0}$ (full black line) is much larger at point B than at point A. 

One would expect that these two effects --- a cavity pull that reduces less with increasing number of photon and a reduced Purcell decay rate --- lead to better qubit measurement at operating point A than B. In the next section, we show numerically that this expectation holds for a Kerr resonator operated close to its bifurcation point. This is done by first calculating the steady-state photon number associated to both qubit states. We then simulate the complete dynamics corresponding to a qubit under measurement with the microwave pulse typically used in bifurcating readouts~~\cite{Siddiqi2006, Lupascu2007, Mallet2009} and which is designed to make the resonator latch in its $H$ state for one of the qubit state. From these simulations, we extract the expected measurement fidelity and show that better results are indeed obtained at operating point A than B. 

\section{Improving bifurcation measurements in the straddling regime} 
\label{sec:improving_bifurcation_measurements_with_the_straddling_regime}
Bifurcation measurements rely on the critical drive amplitude $\epsilon_{H,i}$ --- at which the resonator bifurcate to its high state $H$ --- being different for each qubit state $i$. As illustrated in the inset of Fig.~\ref{fig:bifurcation_phase_diagram_qubit} (c), in bifurcation measurements the measurement drive amplitude $\epsilon_d$ is increased to a value between these two critical amplitudes. However, the bifurcation process being probabilistic, the resonator can still bifurcates from the $L$ to the $H$ state even if the drive amplitude is (slightly) lower than $\epsilon_H$. This yields errors in the measurement and a reduced measurement fidelity. We therefore expect the measurement fidelity to increase with $\Delta\epsilon_H \equiv |\epsilon_{H,0}-\epsilon_{H,1}|$ and so, in other words, with cavity pull. In addition, one expects that a larger separation of the thresholds protects the measurement better against ringing in the resonator's response which, close to $\epsilon_H$, may lead to unwanted bifurcation. For these reasons, we expect that the operating point A, at which the cavity pull should remain larger on a wider range of measurement power, to be better for measurement than point B. 

Below, we first calculate the steady-state response of the resonator in section~\ref{sub:steady_state_response}. We then compute the measurement fidelity for a pulsed measurement in section~\ref{sub:pulsed_measurement_fidelity}. Finally, we discuss other advantages of working in the straddling regime in section~\ref{sub:other_advantages}. 

\subsection{Steady-state response} 
\label{sub:steady_state_response}
We simulate the evolution of the state $\rho$ starting with the resonator in the vacuum and with the qubit either in the eigenstate $\ket{0}$ or $\ket{1}$. We first focus on a drive of constant amplitude $\epsilon_d$, without intrinsic qubit relaxation or dephasing. By looking at the resonator's steady-state response, with this simulation, we want to show that the distance between the bifurcation thresholds $\epsilon_{H,0}$ and $\epsilon_{H,1}$ is indeed larger at operating point A than B. The evolution is governed by the master equation
\begin{equation}
	\label{eqn:master_equation}
	\dot\rho = -i\comm{H}{\rho} + \kappa \sD[a]\rho,
\end{equation}
with the Linblad-form dissipator $\sD[a] = \frac12 (2a\rho\ad - \ada\rho - \rho\ada)$. After a time long compared with $1/\kappa$, we compute the average number of photon $n_i $ for the qubit initially in state $i\elem\{0,1\}$. 

\begin{figure}
	\centering
	\includegraphics[width=0.95\hsize]{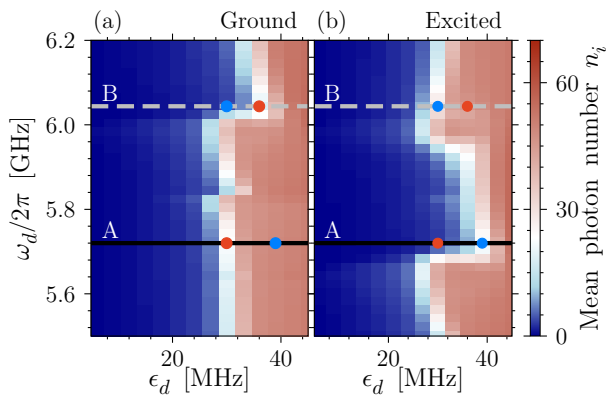}
	\caption{(Color online) Numerically computed average number of photons $n_i$ for a qubit initialized in state $\ket{i}$ with $i=0$ (a) and $i=1$ (b) without qubit relaxation or dephasing. The evolution is computed according to~\eqref{eqn:master_equation}. Qubit parameters are given in the caption of Fig.~\ref{fig:stark_kerr}. Resonator parameters are $(\kappa,K)/2\pi=(5,-0.4)$~MHz and the resonator frequency is adjusted to keep the resonator-drive detuning $(\omega_r-\omega_d)/2\pi=15$~MHz such that the reduced detuning $\Omega/\Omega_C\sim3.5$, well in the bifurcation regime. Dashed lines represent the two operating points A and B (see caption of Fig.~\ref{fig:stark_kerr}). Dots indicates the bifurcation thresholds $\epsilon_{H,i}$ (red for $i=0$, blue for $i=1$). }
	\label{fig:densityplot}
\end{figure}

This quantity is plotted in Fig.~\ref{fig:densityplot} as a function of the drive frequency $\omega_d$ and amplitude $\epsilon_d$ for the qubit initially in its ground (a) or excited (b) state. In both cases, two regions corresponding respectively to the resonator being in the $L$ state (dark blue, $n_i<10$ photons) or in the $H$ state (light red, $n_i\sim50$ photons) can be identified. The border between these two regions (white) corresponds to the critical drive amplitude $\epsilon_{H,i}$, at which the photon population sharply goes from $n_i\sim 15$ to $n_i\sim50$. When comparing these results to the dispersive shifts illustrated in Fig.~\ref{fig:stark_kerr}, we can see that sharp changes in $\varS_i$ and $\varK_i$ translate into sharp changes in the bifurcation amplitudes $\epsilon_{H,i}$. For example, both $\varS_0$ and $\varS_1$ [Fig.~\ref{fig:stark_kerr}~(a)] change sign at $\omega_d/2\pi=6$~GHz, which translate in a sharp change in both $\epsilon_{H,i}$ around that frequency. Moreover, $\varS_1$ changes sign at $\omega_d/2\pi\approx5.7$~GHz, while $\varS_1$ does not. As a result, as can be seen in Fig.~\ref{fig:densityplot}, only $\epsilon_{H,1}$ changes significantly at that frequency. Finally, variations in $\varK_i$ are also visible, for example as the feature in $\epsilon_{H,0}$ at $\omega_d/2\pi\approx5.85$~GHz corresponding to the change of sign in $\varK_0$ at that same frequency. 

The operating points A and B are illustrated in Fig.~\ref{fig:densityplot} by the horizontal full black lines and dashed gray lines respectively. The thresholds $\epsilon_{H,i}$ at these two points are identified by full circles (red for $\epsilon_{H,0}$ and blue for $\epsilon_{H,1}$). As expected from the above arguments, the separation $\Delta\epsilon_H \equiv |\epsilon_{H,0}-\epsilon_{H,1}|$ is larger at A than at B. For the chosen parameters, we find $\Delta\epsilon_H/2\pi \sim10$~MHz at  A while we find $\Delta\epsilon_H/2\pi \sim5$~MHz at B. We note that $\epsilon_{H,1}>\epsilon_{H,0}$ at point A while  $\epsilon_{H,1}<\epsilon_{H,0}$ at point B. This simply changes which resonator state --- of $L$ and $H$ --- is associated with each qubit state.

\subsection{Pulsed measurement fidelity} 
\label{sub:pulsed_measurement_fidelity}
In order to quantify by how much an actual measurement can be improved by working at operating point A --- inside the straddling regime --- rather than at B --- outside of the straddling regime --- we numerically simulated a bifurcation measurement with a sample-and-hold shaped pulse as illustrated in the inset of Fig.~\ref{fig:bifurcation_phase_diagram_qubit}~(c). We recall that, to our knowledge, all experiments with bifurcation measurements have been made outside of the straddling regime so far.

To be more realistic, we performed numerical integration of master equation Eq.~\pref{eqn:master_equation} including qubit dissipation modeled using the Lindblad-form term $\gamma\sum_{i=0}^{M-2}\sD\lsb \frac{g_i}{g_0} \proj{i,i+1}\rsb\rho$. Here, $\gamma$ is the decay rate of the first qubit transition and the factor $g_i/g_0$ is included to take into account the variation of the qubit decay rate with increasing $i$~\cite{Boissonneault2012b}. Pure dephasing is not included since recent devices tend to have very low pure dephasing rates~\cite{Schreier2008,Houck2009}. Including this effect would possibly affect the QND character of the readout due to dressed dephasing~\cite{Boissonneault2008,Boissonneault2009,Wilson2010}, but the extent of this effect has yet to be measured experimentally.

\begin{figure}
	\centering
	\includegraphics[width=0.9\hsize]{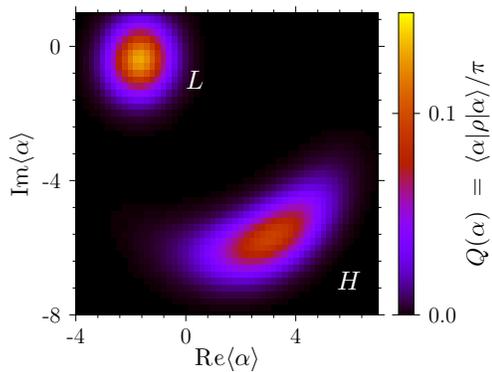}
	\caption{(Color online) Typical $Q$ function $Q(\alpha)$ of a resonator when driven close to its bifurcation threshold $\epsilon_H$.}
	\label{fig:qfunction}
\end{figure}

At the end of the hold time, the $Q$ function of the resonator $Q(\alpha)=\braketop{\alpha}{\rho}{\alpha}/\pi$ is computed. A typical $Q$ function near the bifurcation threshold $\epsilon_H$ is represented in Fig.~\ref{fig:qfunction}. It shows two well-separated smooth peaks corresponding to the $L$ and $H$ states of the resonator. The switching probability is extracted from the weight of the peak that is the farthest away from the origin. From the switching probabilities, the worst-case error probability
\begin{equation}
	P_{\rm error} = \max_{\substack{\{i,j\}\elem\{0,1\} \\ j\ne i}} P(j|i),
\end{equation}
can be computed and where $P(j|i)$ is the probability of assigning the measurement to the qubit state $\ket{j}$, given that the qubit was initially in $\ket{i}$. This numerical procedure was previously tested against experimental single-shot bifurcation measurement of a transmon qubit~\cite{Mallet2009} and found an identical measurement fidelity, within a margin of 2\%~\cite{Bertet2011}. 

\begin{figure}
	\centering
	\includegraphics[width=0.95\hsize]{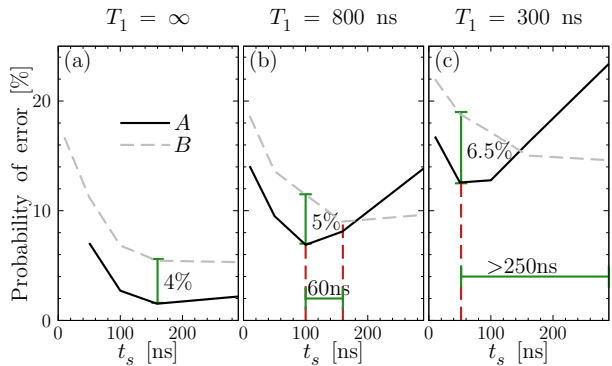}
	\caption{(Color online) Error probability for the outcome of a bifurcation measurement versus sampling time $t_s$. The measurement pulsed is illustrated in the inset of Fig.~\ref{fig:bifurcation_phase_diagram_qubit}~(c). Full black lines correspond to operating point A, in the straddling regime and dashed grey lines correspond to operating point B, outside of the straddling regime. The parameters are the same as in Fig.~\ref{fig:densityplot}. The simulations are realized for a qubit with intrinsic relaxation times (a) $T_1=\inf$, (b) $800$~ns and (c) $300$~ns. In (b) and (c), the dashed vertical red lines indicate the minimum of the curves for the two operating points. Full green bars indicate the gain in measurement time (horizontal) or measurement fidelity (vertical) between the two operating points.}
	\label{fig:prob_error}
\end{figure}

We show in Fig.~\ref{fig:prob_error} the worst-case error probability as a function of the sampling time $t_s$ for three different qubit relaxation times $T_1$. These results have been obtained by minimizing the error probability with respect to  $\sigma$ and $\delta\epsilon_d$ [see inset of Fig~\ref{fig:bifurcation_phase_diagram_qubit}~(c) for definitions]. Comparing panels (a), (b) and (c), we see that $P_{\rm error}$ increases as the qubit relaxation time decreases, which is expected because of the increased odds of the qubit relaxing before the resonator switches from $L$ to $H$.

We now compare the results inside (full black lines, operating point A) and outside (dashed grey lines, operating point B) of the straddling regime. We first observe that for short sampling times $t_s$, the error probability is always lower for operating point A than B. Since the low-photon cavity pulls $\chi$ were chosen to be the same for both points, this improvement is due both to the sign and amplitude of the Kerr terms $\varK_i$ and to the reduced Purcell decay as explained in section~\ref{sec:dispersive_model_in_the_straddling_regime}. The situation is however reversed for larger $t_s$ where point B is superior.
As illustrated in Fig.~\ref{fig:densityplot}, this is because the resonator switches at a lower power for the ground state than for the excited state at point A, while the opposite is true for point B. This implies that qubit relaxation induces resonator switching (ie false positives) at point A, but not at point B. We note that the situation would be reversed for a qubit with a positive anharmonicity such as the low-impedance flux qubit~\cite{Steffen2010}, increasing further the advantage of working in the straddling regime.

Overall, we find that operating within the straddling regime always allows to reach lower error probabilities with a sampling time $t_s$ always as short, or shorter, than outside of the straddling regime. When operating in the straddling regime, the error probability is up to 3 times smaller than outside. Finally, the absolute improvement is better for qubits with shorter lifetimes, but as expected the best fidelity is found for qubits with longer lifetimes. 


\subsection{Other advantages} 
\label{sub:other_advantages}
The above improvement in readout fidelity has been obtained by working with a qubit-resonator coupling $g$ that is more than an order of magnitude smaller than current experimental realizations. Lower coupling however leads to slower two-qubit gates when these rely on qubit-qubit interactions mediated by the same resonator mode that is used for readout. This problem can be sidestepped by either taking advantage of different modes for readout and two-qubit gates~\cite{Leek2010} or, as recently experimentally realized, using direct capacitive coupling between the qubits~\cite{Dewes2011,Dewes2011a}.

With the above problem avoided, working with weaker coupling $g$ can be advantageous in other ways than the more efficient readout studied here. For example, it allows to greatly reduce Purcell decay by biasing the qubit away from a resonator resonance when it is not being measured. With a reduction by a factor of 10 of the coupling, a reduction by a factor of 100 of the Purcell decay rate can be obtained for the same detuning and cavity damping [see full black line in Fig.~\ref{fig:stark_kerr}(e)]. At the time of measurement, the qubit-resonator detuning can be adjusted such as to reach the straddling regime. This can be done by changing the flux in the qubit loop or by using a tunable resonator (or both)~\cite{Wallquist2006}. Moving in and out of the straddling regime in this way necessarily means going through a qubit-resonator crossing. With a large coupling $g$, the associated (and 
unwanted) Landau-Zener-Stueckelberg transitions can be correspondingly large~\cite{Zueco2008}. This probably is however greatly reduced when working with small couplings. Indeed, assuming a frequency-tuning speed of $v=2\pi \cdot 1~{\rm GHz}/1~{\rm ns}$, one finds the probability of unwanted transition $P=1-e^{-2\pi g_{0}^2/v}=0.7\%$ for a coupling $g/2\pi= 13.5$~MHz, while the same probability is $\sim10\%$ for $g/2\pi = 50$~MHz and $\sim30\%$ for $g/2\pi= 100$~MHz.

Smaller coupling strengths can also help in reducing spectral crowding in the presence of multiple qubits coupled to a single resonator. Indeed, even if the qubit-qubit interaction mediated by virtual excitations of the resonator is not actively used for logical gates, it is always present and can lead to errors. The rate of this interaction can be reduced by increasing the qubit-qubit detuning by an amount that is large with respect to the coupling $g$. With large $g$ and multiple qubit, the available spectral range (typically from $\sim$ 4 to 15 GHz) is rapidly occupied and only a few qubits can be coupled to the same resonator without having to deal with unwanted two-qubit gates. Using the straddling regime to increase the measurement fidelity with smaller coupling addresses this problem and does not require advanced circuit designs~\cite{Mariantoni2011a}. 

\section{Conclusion} 
\label{sec:conclusion}
We have studied the measurement of a multi-level superconducting qubit using bifurcation of a Kerr nonlinear resonator and by exploiting the straddling regime. The method is applicable to any qubit with a weakly anharmonic multi-level structure with only nearest-level transitions, but could be generalized to more complex structures and couplings. As we have shown, working in the straddling regimes allows larger qubit-state-dependent pulls of the resonator frequency for a given coupling or, equivalently, the same pull for smaller couplings. While outside of the straddling regime, the resonator frequency shift is reduced at higher photon numbers~\cite{Boissonneault2008}, we show that, inside the straddling regime, it is possible to find operating points where this reduction is minimized. We also show that the Purcell decay rate can be much smaller for a given cavity pull inside the straddling regime. Combined, these two effects lead to an increased fidelity for bifurcation measurements and we find an error probability up to three times smaller inside than outside of the straddling regime for a sampling time that can be more than $250$~ns shorter. We find measurement fidelities $1-P_{\rm error}$ larger than 98\% with a qubit-resonator coupling as small as $13.5$~MHz with realistic system parameters.

The method presented in this paper has also the advantage of reducing spectral crowding in multiple-qubit systems. It does that without requiring complex circuits and allows to effectively remove Purcell decay when the qubits are not being measured.

\begin{acknowledgments}
AB acknowledges funding from NSERC, the Alfred P. Sloan Foundation, and CIFAR. MB acknowledges funding from NSERC and FQRNT. We thank Calcul Qu\'ebec and Compute Canada for computational resources.
\end{acknowledgments}

\appendix

\section{Dispersive transformation of the two-photon terms} 
\label{annsec:dispersive_transformation_of_the_two_photon_terms}
In this Appendix, we follow Ref.~\cite{Boissonneault2012b} to diagonalize the Hamiltonian~\pref{eqn:H} as well as a two-photon transition term that can be large only in the straddling regime. To do so, we first apply a polaron transformation~\cite{Mahan2000,Irish2005,Gambetta2008}
\begin{equation}
	\label{eqn:polaron_transformation}
	\tP = \sum_{i=0}^{M-1} \proj{i,i} D(\alpha_i),
\end{equation}
where $D(\alpha)$ is a displacement transformation~\cite{Scully1997}
\begin{equation}
	\label{eqn:displacement_transformation}
	D(\alpha) = \exp\lsb \alpha\ad - \alpha^* a\rsb,
\end{equation}
that displaces the resonator field operator $a\rightarrow a+\alpha_i$. The result of the polaron transformation on $a$ is therefore $a\rightarrow a+\proj{\alpha}$, where $\proj{\alpha}$ is defined according to \eqref{eqn:proj_x}. We follow this polaron transformation by a dispersive transformation of the classical detuned drive on the qubit
\begin{equation}
	\label{eqn:classical_dispersive_transformation}
	\tD_C = \exp\lsb \sum_{i=0}^{M-2} \xi_i^* \proj{i,i+1} - \xi_i \proj{i+1,i}\rsb,
\end{equation}
where $\xi_i$ is a classical analogue of the operator $\lambda_i \ad$ in the dispersive transformation~\cite{Carbonaro1979,Boissonneault2008}. Applying these two transformations on the Hamiltonian~\pref{eqn:H} and choosing $\alpha_i$ according to \eqref{eqn:condition_alpha_d} and
\begin{equation}
	\label{eqn:xi_i}
	\xi_{i} = \frac{-g_i\alpha_i}{\omega_{i+1,i}-\omega_d},
\end{equation}
yields the Hamiltonian~\cite{Boissonneault2012b}
\begin{equation}
	\label{eqn:H_second}
	\begin{split}
		H'' &= \sum_{i=0}^{M-1} \omega_i'' \proj{i,i} + H_I + \omega_r'(\alpha)\ada \\
		&\quad + \sum_{i=0}^{M-3} \alpha_i^*\alpha_{i+1}^* e^{-2i\omega_d t}g^{(2)}_i \proj{i,i+2} + \hc,
	\end{split}
\end{equation}
where the dispersive transformation has been performed to fourth order and $g_i^{(2)}$, given at \eqref{eqn:g_i_2}, is an effective coupling due to two-photon transitions. In the above Hamiltonian, we have defined the ac-Stark shifted qubit frequencies
\begin{equation}
	\label{eqn:omega_second}
	\omega_i'' = \omega_i + \varS_i |\bar\alpha|^2 + \varK_i^{(1)}|\bar\alpha|^4,
\end{equation}
where $\bar\alpha=\mean{\proj{\alpha}}$, $\alpha_i$ is given by the solution of \eqref{eqn:condition_alpha_d}, $\varS_i$ is given at \eqref{eqn:varS_i}, while $\varK_i^{(1)}$ is given by the first three lines of \eqref{eqn:varK_i}, and the Kerr-shifted resonator frequency
\begin{equation}
	\label{eqn:omega_r_prime}
	\omega_r'(\alpha) \equiv \omega_r+2K|\bar\alpha|^2.
\end{equation}

We note that the second line of $H''$ is not diagonal. In Ref.~\cite{Boissonneault2012b}, this term was dropped assuming that $g_i^{(2)}$ was small and that $|2\omega_d - \omega_{i+2}+\omega_i|$ was large enough to do a RWA. Here however, since we are interrested in the straddling regime, the same can not be done. Indeed, if, for example, the drive frequency is $\omega_d = (\omega_2-\omega_1)/2 + (\omega_1-\omega_0)/2$, which falls directly in the middle of a straddling regime, the second line of $H''$ is resonant and a two-photon transition from $\ket{0}$ to $\ket{1}$ is driven. Moreover, since $\Delta_{i+1,d}$ and $\Delta_{i,d}$ have the same sign, the coupling $g_i^{(2)}$ can be large. We can however approximately diagonalize this term using a second transformation of the form
\begin{equation}
	\tD^{(2)} = \exp\lsb \sum_{i=0}^{M-3} {\xi_i^{(2)}}^* \proj{i,i+2} - \xi_i^{(2)} \proj{i+2,i}\rsb.
\end{equation}
Applying this transformation on \eqref{eqn:H_second} and choosing
\begin{equation}
	\label{eqn:xi_i2}
	\xi_i^{(2)} = \frac{-g_i^{(2)} \alpha_i\alpha_{i+1}}{\Delta_{i+1,d}+\Delta_{i,d}},
\end{equation}
yields a correction to the Kerr shift, giving \eqref{eqn:varK_i}. Applying a final dispersive transformation on $H''$ in order to diagonalize the quantum interaction $H_I$ yields the diagonalized Hamiltonian~\pref{eqn:H_fourth}.


\bibliography{/Users/mboisson/Documents/Articles/Biblio.bib}

\end{document}